\documentclass[aps,prl,superscriptaddress,showpacs,preprintnumbers,byrevtex,amsmath,amssymb]{revtex4}

\usepackage{graphicx} 
\usepackage{dcolumn}  

\newcommand{\ra}{\ensuremath{\rightarrow}}
\newcommand{\re}{\ensuremath{(0.033\,^{+0.007}_{-0.006} \pm 0.009)}}
\newcommand{\rd}{\ensuremath{(0.87\,^{+0.32}_{-0.28} \pm 0.20)}}
\newcommand{\rs}{\ensuremath{(0.53\,^{+0.19}_{-0.15} \pm 0.14)}}
\newcommand{\rdcc}{\ensuremath{(0.74\,^{+0.28}_{-0.24} \pm 0.19)}}
\newcommand{\rscc}{\ensuremath{(1.01\,^{+0.36}_{-0.30} \pm 0.26)}}
\newcommand{\racc}{\ensuremath{(0.87\,^{+0.21}_{-0.19} \pm 0.17)}}
\newcommand{\raccrel}{\ensuremath{0.59\,^{+0.15}_{-0.13} \pm 0.12}}

\begin{document}

\vspace*{-3\baselineskip}
\resizebox{!}{3cm}{\includegraphics{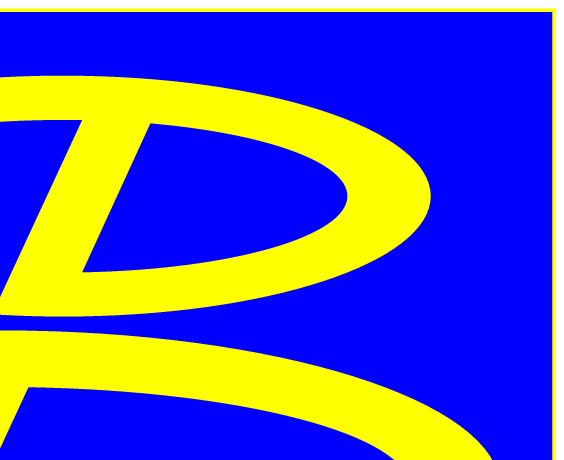}}


\preprint{\vbox{ \hbox{Belle Preprint 2002-13}
                 \hbox{KEK Preprint 2002-27}}}

\title{ \quad\\[0.5cm]  Observation of double $c{\bar{c}}$
production in $e^{+}e^{-}$ annihilation at $\sqrt{s} \approx
10.6~$GeV}

\affiliation{Aomori University, Aomori}
\affiliation{Budker Institute of Nuclear Physics, Novosibirsk}
\affiliation{Chiba University, Chiba}
\affiliation{Chuo University, Tokyo}
\affiliation{University of Cincinnati, Cincinnati OH}
\affiliation{Deutsches Elektronen--Synchrotron, Hamburg}
\affiliation{University of Frankfurt, Frankfurt}
\affiliation{Gyeongsang National University, Chinju}
\affiliation{University of Hawaii, Honolulu HI}
\affiliation{High Energy Accelerator Research Organization (KEK), Tsukuba}
\affiliation{Hiroshima Institute of Technology, Hiroshima}
\affiliation{Institute of High Energy Physics, Chinese Academy of Sciences, Beijing}
\affiliation{Institute of High Energy Physics, Vienna}
\affiliation{Institute for Theoretical and Experimental Physics, Moscow}
\affiliation{J. Stefan Institute, Ljubljana}
\affiliation{Kanagawa University, Yokohama}
\affiliation{Korea University, Seoul}
\affiliation{Kyoto University, Kyoto}
\affiliation{Kyungpook National University, Taegu}
\affiliation{Institut de Physique des Hautes \'Energies, Universit\'e de Lausanne, Lausanne}
\affiliation{University of Maribor, Maribor}
\affiliation{University of Melbourne, Victoria}
\affiliation{Nagoya University, Nagoya}
\affiliation{Nara Women's University, Nara}
\affiliation{National Kaohsiung Normal University, Kaohsiung}
\affiliation{National Lien-Ho Institute of Technology, Miao Li}
\affiliation{National Taiwan University, Taipei}
\affiliation{H. Niewodniczanski Institute of Nuclear Physics, Krakow}
\affiliation{Nihon Dental College, Niigata}
\affiliation{Niigata University, Niigata}
\affiliation{Osaka City University, Osaka}
\affiliation{Osaka University, Osaka}
\affiliation{Panjab University, Chandigarh}
\affiliation{Peking University, Beijing}
\affiliation{Princeton University, Princeton NJ}
\affiliation{RIKEN BNL Research Center, Brookhaven NY}
\affiliation{University of Science and Technology of China, Hefei}
\affiliation{Seoul National University, Seoul}
\affiliation{Sungkyunkwan University, Suwon}
\affiliation{University of Sydney, Sydney NSW}
\affiliation{Tata Institute of Fundamental Research, Bombay}
\affiliation{Toho University, Funabashi}
\affiliation{Tohoku Gakuin University, Tagajo}
\affiliation{Tohoku University, Sendai}
\affiliation{University of Tokyo, Tokyo}
\affiliation{Tokyo Institute of Technology, Tokyo}
\affiliation{Tokyo Metropolitan University, Tokyo}
\affiliation{Tokyo University of Agriculture and Technology, Tokyo}
\affiliation{Toyama National College of Maritime Technology, Toyama}
\affiliation{University of Tsukuba, Tsukuba}
\affiliation{Utkal University, Bhubaneswer}
\affiliation{Virginia Polytechnic Institute and State University, Blacksburg VA}
\affiliation{Yokkaichi University, Yokkaichi}
\affiliation{Yonsei University, Seoul}

\author{K.~Abe}               
\affiliation{High Energy Accelerator Research Organization (KEK), Tsukuba}
\author{K.~Abe}               
\affiliation{Tohoku Gakuin University, Tagajo}
\author{R.~Abe}               
\affiliation{Niigata University, Niigata}
\author{T.~Abe}               
\affiliation{Tohoku University, Sendai}
\author{I.~Adachi}            
\affiliation{High Energy Accelerator Research Organization (KEK), Tsukuba}
\author{Byoung~Sup~Ahn}       
\affiliation{Korea University, Seoul}
\author{H.~Aihara}            
\affiliation{University of Tokyo, Tokyo}
\author{M.~Akatsu}            
\affiliation{Nagoya University, Nagoya}
\author{Y.~Asano}             
\affiliation{University of Tsukuba, Tsukuba}
\author{T.~Aso}               
\affiliation{Toyama National College of Maritime Technology, Toyama}
% ...

\collaboration{Belle Collaboration}
\noaffiliation

\begin{abstract}
We report the observation of prompt $J/\psi$ via double $c{\bar{c}}$
production from the $e^{+}e^{-}$ continuum. In this process one
$c{\bar{c}}$ pair fragments into a $J/\psi$ meson while the remaining pair 
either produces a bound charmonium state or fragments into open charm.
Both cases have been observed: the first by studying
the mass spectrum of the system recoiling against the $J/\psi$,
and the second by reconstructing the $J/\psi$ together with a charmed meson.
We find cross-sections of
$\sigma (e^{+}e^{-} \ra J/\psi\, \eta_c (\gamma)) \times
{\mathcal{B}}(\eta_c \ra \,\geq 4\,{\rm charged}) = \re$ pb and
$\sigma (e^{+}e^{-} \ra J/\psi\, D^{\ast +} X) = \rs$ pb,
and infer
$\sigma(e^{+}e^{-} \ra J/\psi\,c\bar{c}) /
 \sigma(e^{+}e^{-} \ra J/\psi\,X) = \raccrel$;
in each case the uncertainty quoted second is the systematic error. 
These results are obtained from a $46.2\,{\rm
fb}^{-1}$ data sample collected near the $\Upsilon(4S)$ resonance,
with the Belle detector at the KEKB asymmetric energy $e^+ e^-$ collider.
\end{abstract}


\maketitle

\tighten

\newpage

{\renewcommand{\thefootnote}{\fnsymbol{footnote}}}
\setcounter{footnote}{0}

Prompt charmonium production in $e^+e^-$ annihilation provides an
opportunity to study both perturbative and non-perturbative effects in
QCD. In a previous paper~\cite{behera} we presented cross-section
measurements for prompt $J/\psi$ and $\psi(2S)$ production, and studies of their
kinematic properties, which were compared to predictions of the
NRQCD~\cite{nrqcd1,nrqcd2a,nrqcd2c,nrqcd2d} model.
The BaBar Collaboration has also
published results on prompt $J/\psi$ production~\cite{babar}.
NRQCD predicts that prompt $J/\psi$ production at
$\sqrt{s} \approx 10.6\,\mathrm{GeV}$
is dominated by $e^+ e^- \ra J/\psi\, gg$, with additional contributions
from $J/\psi\, g$, $J/\psi\, c\bar{c}$ and other processes.
The color-octet $J/\psi\, g$ signal predicted in Refs~\cite{nrqcd2c,nrqcd2d}
is not observed~\cite{behera}.
The results in Refs~\cite{behera,babar} do not constrain the contributions 
from $J/\psi\, gg$ and $J/\psi\, c\bar{c}$.

We present the results of a search for $J/\psi\,
c\bar{c}$ production, \emph{i.e.}\ double $c\bar{c}$ production,
where the additional $c\bar{c}$ pair fragments into either charmonium
or charmed hadrons.
Exclusive associated production with a $J/\psi$ meson was proposed in Ref.~\cite{iwa}
as a method for searching for the remaining undiscovered charmonium states.
Based on a non-perturbative approach the model of Ref.~\cite{kane} predicts a
cross-section of a few pb for $J/\psi\,\eta_c$ production at
$\sqrt{s}\approx 10.6\,\mathrm{GeV}$. In contrast, recent theoretical
estimates of the $J/\psi\, c\bar{c}$ cross-section are 
as small as $0.07\,{\mathrm{pb}}$~\cite{nrqcd2a,nrqcd2d,lik}.

This analysis is based on $41.8\,\mathrm{ fb}^{-1}$ of data at the
$\Upsilon(4S)$ and $4.4\,\mathrm{fb}^{-1}$ at an energy 
$60\,{\mathrm{MeV}}$ below the resonance, collected with the Belle detector
at the KEKB asymmetric energy storage rings~\cite{KEKB}.
Belle is a large solid-angle magnetic spectrometer~\cite{Belle}.
Charged particles are reconstructed in a
50-layer central drift chamber (CDC) and their impact parameters are
determined using a three-layer silicon vertex detector (SVD).
Kaon/pion separation is based on specific ionization ($dE/dx$)
in the CDC, time of flight measurements and the response
of aerogel \v Cerenkov counters (ACC). Electron identification is
based on a combination of $dE/dx$ measurements, the ACC response and
information about the shape, energy deposit and position of the
associated shower in the electromagnetic calorimeter (ECL). Muon
identification is provided by 14 layers of $4.7\,\mathrm{cm}$ thick
iron plates interleaved with resistive plate counters.
Photons are reconstructed in the ECL as showers with an energy
larger than $20\,\mathrm{MeV}$ that are not associated with charged
tracks.

We use hadronic events separated from QED, $\tau\tau$,
two-photon and beam-gas interaction backgrounds by the 
selection criteria described in Ref.~\cite{behera}.  Charged pion and kaon
candidates are well reconstructed tracks that have been 
positively identified.  $K_S$ candidates are reconstructed by
combining $\pi^+ \pi^-$ pairs with an invariant mass within
$10\,\mathrm{MeV}/c^2$ of the nominal $K_S$ mass.
We require the distance between the pion tracks at the $K_S$ vertex
to be  less than $1\,\mathrm{cm}$,
the transverse flight distance from the interaction point
to be greater than $0.5\,\mathrm{cm}$
and the angle between the $K_S$ momentum direction and decay path
to be smaller than $0.1\,\mathrm{rad}$.
Photons of energy greater than $50\,\mathrm{MeV}$ are combined to form
$\pi^0 \rightarrow \gamma\gamma$
candidates if their invariant mass lies within $10\,\mathrm{ MeV}/c^2$
of the nominal $\pi^0$ mass. Such $\gamma\gamma$ pairs are fitted with a
$\pi^0$ mass constraint to improve the momentum resolution.

The $J/\psi \ra \ell^{+}\ell^{-}$ reconstruction procedure is identical
to that presented in Ref.~\cite{behera}. Two positively identified lepton
candidates are required to form a common vertex that is less than 
$500\,\mu\mathrm{m}$ from the interaction point in the plane
perpendicular to the beam axis. For $J/\psi \ra e^{+}e^{-}$, the invariant mass
calculation includes the four-momentum of photons detected within
$50\,\mathrm{mrad}$ of the $e^{\pm}$ directions, as a partial
correction for final state radiation and bremsstrahlung energy loss.
QED processes are suppressed by requiring the total charged
multiplicity ($N_{\rm ch}$) to be greater than four. $J/\psi$
mesons from $B\overline{B}$ events are removed by requiring
the momentum of the meson in the $e^+e^-$ center of mass (CM) system,
$p^{*}_{J/\psi}$, to be greater than $2.0\,\mathrm{GeV}/c$. The process
$e^{+}e^{-} \ra \psi(2S) \gamma $ is partially removed by rejecting
events with a photon of CM energy $E^\ast > 3.5\,\mathrm{GeV}$.

The $J/\psi \ra \ell^{+}\ell^{-}$ signal region is defined by the mass
window $ \left|M_{\ell^{+}\ell^{-}}-M_{J/\psi}\right| < 30 \, \mathrm{
MeV}/c^2$ ($\approx 2.5\, \sigma$). The sideband region is defined by
$ 100 < \left| M_{\ell^{+}\ell^{-}} -M_{J/\psi} \right| <
400\,\mathrm{MeV}/c^2$ and is used to estimate the contribution from
the dilepton combinatorial background under the $J/\psi$ peak.
The sideband is scaled, parameterizing the background by a quadratic function.

\begin{figure}[htb]
\includegraphics[width=0.48\textwidth]{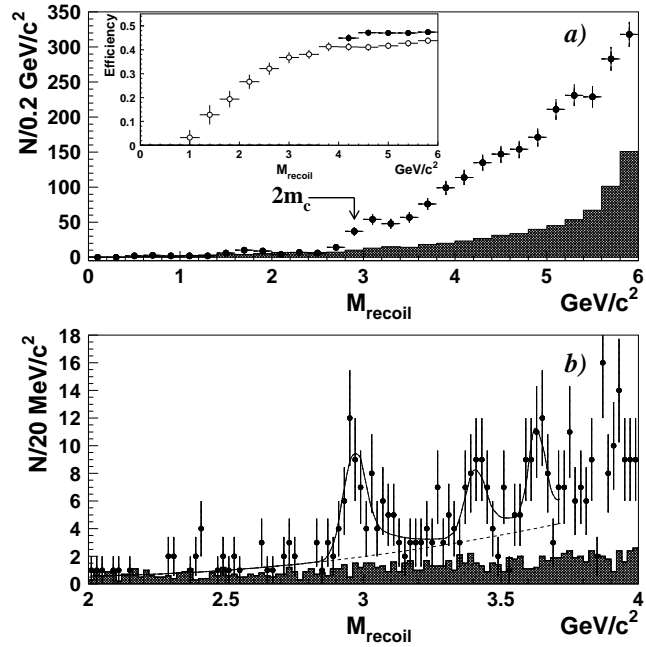}
\caption{a) The recoil mass distribution for the $J/\psi$ signal region
(points) and scaled sidebands (shaded histogram).
The inset shows the MC reconstruction efficiency for
$e^+e^- \ra J/\psi\,q\bar{q}$ ($q = u,d,s$; open circles) and
$e^+e^- \ra J/\psi\,c\bar{c}$ events (closed circles).
b)
The recoil mass distribution after refitting the $J/\psi$ candidate with a
mass constraint for the $J/\psi$ signal region (points)
and scaled sidebands (shaded histogram). The curve represents
the fit described in the text.}
\label{xm}
\end{figure}

Distributions of the mass of the system recoiling against the $J/\psi$
candidate---the ``recoil mass''---after this selection 
are shown in Fig.~\ref{xm}{\it a}, for both signal and sideband regions.
The recoil mass is defined as
\begin{equation}
M_{\rm recoil} = \sqrt{\left(\sqrt{s}-E_{J/\psi}^*\right)^2-p_{J/\psi}^{*~2}}.
\end{equation}
A clear threshold near $2\, m_c$ can be seen.
To avoid model dependence, we do not perform an acceptance correction.
As examples, the Monte Carlo efficiencies for both
$e^+e^- \ra J/\psi\,q\bar{q}$ ($q = u,d,s$)
and $J/\psi\,c\bar{c}$ events are shown in the inset:
the variation with $M_{\rm recoil}$ is smooth. 
For recoil masses smaller than $2.8\,\mathrm{GeV}/c^2$, no
significant $J/\psi$ signal is observed: a fit to the dilepton mass 
spectrum for events in this region gives $16.4 \pm 7.8$ events.  

The region $2\, m_c \lesssim M_{\rm recoil} < 2\, m_D$ is
studied in more detail in order to search for production of $J/\psi$ together
with an additional charmonium state.
We apply a mass constrained fit to the $J/\psi$ candidates before
determining $p_{J/\psi}^*$. Monte Carlo simulation predicts that this
improves the $M_{\rm recoil}$ resolution by a factor close
to two. The resulting recoil mass spectrum in the data is presented in
Fig.~\ref{xm}{\it b}: a clear peak is observed around $3 \,
\mathrm{GeV}/c^2$. Since $e^{+}e^{-} \ra \gamma^* \ra J/\psi\, J/\psi$ 
is forbidden by charge conjugation symmetry, we interpret this peak as
$e^{+}e^{-} \ra J/\psi\, \eta_c$.
Additional peaks at recoil masses consistent with the 
$\chi_{c0}$ and $\eta_c(2S)$ mass~\cite{etaprimemass} are also seen.

To reproduce the $\eta_c$ shape in the recoil mass spectrum we
generate $e^{+}e^{-} \ra \gamma^*\, (\gamma) \ra J/\psi\, \eta_c\,
(\gamma)$ Monte Carlo events.
We assume $p$-wave $J/\psi\, \eta_c$ production, as required by
parity conservation, and phase space suppression of 
$\gamma^\ast \ra J/\psi\, \eta_c$ as the virtual $\gamma^\ast$ energy
varies due to initial state radiation (ISR). This effect produces
a high-mass tail in $M_{\rm recoil}$. The corresponding
$\eta_c(2S)$ lineshape is narrower, due to the
larger mass of the state and its presumed smaller intrinsic width.
The $\chi_{c0}$ lineshape assumes $s$-wave production.
The effect of varying the energy dependence of the cross-section in
the ISR calculation is included in the systematic
error.

A fit to the recoil mass
spectrum finds an $\eta_c$ yield $N_{\eta_c}=67^{+13}_{-12}$ at a mass
$M=(2.962 \pm 0.013)\,\mathrm{GeV}/c^2$, with $\chi_{c0},
\eta_c(2S)$ yields $N_{\chi_{c0}} = 39^{+14}_{-13}$,
$N_{\eta_c(2S)} = 42^{+15}_{-13}$ at masses $M_{\chi_{c0}} =
(3.403\pm 0.014)\,\mathrm{GeV}/c^2$ and $M_{\eta_c(2S)} = (3.622\pm
0.012)\,\mathrm{GeV}/c^2$ respectively. We use a second order
polynomial to describe the background.
Only the region below the open
charm threshold ($M_{\rm recoil}<3.73\,\mathrm{GeV}/c^2$) is included
in the fit.

We assess the significance of each signal $i$ using 
$\sigma_i \equiv \sqrt{-2\ln(\mathcal{L}_0^i / \mathcal{L}_{max})}$,
where $\mathcal{L}_{max}$ is the maximum likelihood returned by the
fit, and $\mathcal{L}_0^i$ is the likelihood with the yield of the
state $i$ ($i = \eta_c, \chi_{c0}, \eta_c(2S))$ set to zero. We
find $\sigma_{\eta_c} = 6.7$, $\sigma_{\chi_{c0}} = 3.3$ and
$\sigma_{\eta_c(2S)} = 3.4$.  As the significance of the
$\chi_{c0}$ and $\eta_c(2S)$ peaks is low, we perform an additional
fit using only the $\eta_c$ shape and the polynomial
background: this finds $N_{\eta_c}= 56^{+13}_{-12}$ with 
$\sigma_{\eta_c} = 5.9$.
We use this result,
together with the results of fits after varying the
charmonium intrinsic widths~\cite{PDG} and the choice of background shape,
to estimate the systematic error on the $\eta_c$ yield
due to the fitting procedure.

To determine the $e^{+}e^{-} \ra J/\psi\, \eta_c\, (\gamma)$
cross-section we correct the signal yield for the
reconstruction efficiency obtained from the Monte Carlo.
Because of the requirement $N_{\rm ch}>4$, the recoil system must
contain at least three charged tracks:
this removes $\eta_c$ decays into 0 or 2 charged tracks plus neutrals.
As $\eta_c$ branching fractions are poorly known, we express our
result in terms of the product $\sigma (e^{+}e^{-} \ra J/\psi\,
\eta_c\, (\gamma)) \times {\mathcal{B}}(\eta_c \ra \,\geq 4\,{\rm
charged})$, which we find to be \re$\,{\mathrm {pb}}$.
Here and elsewhere, the uncertainty quoted second is the systematic error.
The various sources of systematic error are listed in Table~\ref{sys1},
with dominant contributions from the uncertainty in the ISR correction,
the choice of fitting procedure, and the production and helicity angle
distributions for the $J/\psi$ meson.
\begin{table}[htb]
\caption{Sources of systematic error for $\sigma(e^+ e^- \to J/\psi\, \eta_c$). }
\label{sys1}
\begin{tabular}
{@{\hspace{0.5cm}}l@{\hspace{0.5cm}}||@{\hspace{0.5cm}}c@{\hspace{0.5cm}}}
\hline \hline
Source & Systematic error (\%) \\
\hline
ISR correction & $\pm 19$ \\
Fitting procedure & $\pm 16$ \\
$J/\psi$ polarization & $\pm 11$ \\
Track reconstruction  & $\pm 5$ \\
Lepton identification  & $\pm 4$ \\
\hline
Total & $\pm 28$ \\
\hline \hline
\end{tabular}
\end{table}

To study the $J/\psi\, c\bar{c}$ mechanism in the region $M_{\rm
recoil} \geq 2\, m_{D}$, we search for fully reconstructed $D^{\ast +}$
and $D^0$ decays~\cite{implied} in events with a $J/\psi$ meson.
For the study of $J/\psi\,
D^{\ast +}$ associated production we reconstruct $D^{\ast +} \ra D^0\,
\pi^+$ using five $D^0$ decay modes: $K^- \pi^+$, $K^- K^+$,
$K^- \pi^- \pi^+ \pi^+$, $K_S \pi^+ \pi^-$ and $K^- \pi^+ \pi^0$.  We
select $D^0$ candidates in a $\pm 10\,\mathrm{MeV}/c^2$ mass window
for the charged modes and a $\pm 20\,\mathrm{MeV}/c^2$ window for
$K^- \pi^+ \pi^0$ (approximately $2\,\sigma$ in each case).
To improve the $M_{D^0 \pi^+}$ resolution $D^0$ candidates are
refitted to the nominal $D^0$ mass.

Although $B \to J/\psi\,X$ decays are rejected by the
selection, semileptonic $B$ decays contribute to the
background under the $J/\psi$ peak and lead to a large $D^{\ast +}$
signal. To remove the remaining $B\overline{B}$ background we require
that either the $D^{\ast +}$ or one of the leptons from the $J/\psi$
candidate have a momentum above the kinematic limit for $B$ decays:
$p^*_{D^{\ast +}}>2.6\,\mathrm{GeV}/c$ or
$p^*_{\ell^\pm}>2.6\,\mathrm{GeV}/c$.
Choosing the $D^0 \pi^+$ combination with the best $D^0$ mass yields at
most one $J/\psi\,D^{\ast +}$ candidate per event.

\begin{figure}[htb]
\includegraphics[width=0.48\textwidth]{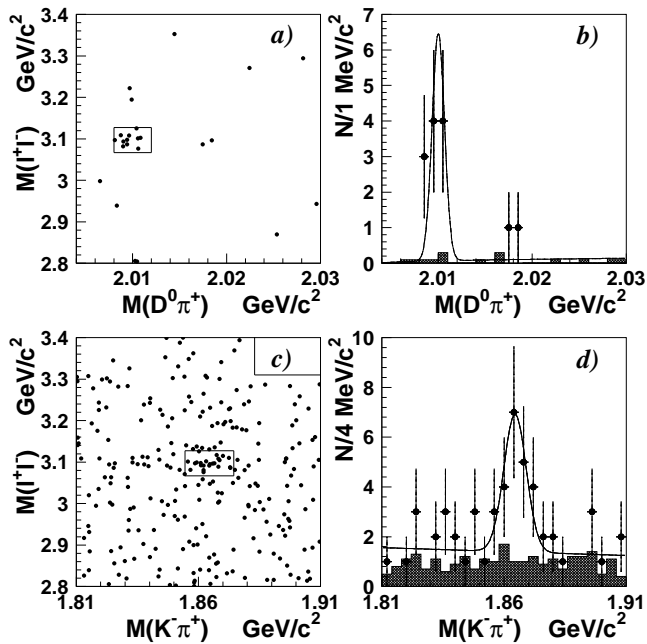}
\caption{Results of a search for associated production of $J/\psi$ and
charm mesons: a) the scatter plot $M(l^{+}l^{-})$ vs $M(D^0 \pi^+)$;
b) projection onto the $M(D^0 \pi^+)$ axis; c) the scatter plot
$M(l^{+}l^{-})$ vs $M(K^- \pi^+ (K^+ K^-))$; d) projection onto the
$M(K^- \pi^+ (K^+ K^-))$ axis. Points with error bars show the $J/\psi$ signal
region and the shaded histograms show the scaled sidebands. The curves
represent the fit described in the text.}
\label{md} 
\end{figure}

The scatter plot of the dilepton mass versus the $D^0 \pi^+$ mass, and
the $D^0 \pi^+$ mass projection, are shown in Figs.~\ref{md}{\it
a,b}.
We perform a fit to the $D^0 \pi^+$ mass distribution in the $J/\psi$ signal
window, with a Gaussian for the $D^{\ast +}$ signal and a threshold function
$A\sqrt{M_{D^0 \pi^+}-M_{\rm {thres}}}$ for the background.
The $J/\psi$ sideband is fit simultaneously, and used to estimate the 
combinatorial $(\ell^+\ell^-)\,D^{\ast +}$ contribution to the $D^{\ast +}$
yield in the signal region. We find $N_{D^{\ast +}}=10.5^{+3.6}_{-3.0}$,
with a combinatorial contribution of $0.4\pm 0.3$: the signal yield is
$10.1^{+3.6}_{-3.0}$, with significance $\sigma_{J/\psi\,D^{\ast +}\,X} = 5.3$.
As a cross-check we fit the dilepton mass distribution in the region
$2.008<M_{D^0 \pi^+}<2.012\,\mathrm{GeV}/c^2$, finding
$N_{J/\psi}=9.6^{+3.6}_{-2.9}$. The $J/\psi$ signal shape
is fixed from the Monte Carlo simulation and we include a linear
background function.

For the study of $J/\psi\, D^0$ associated production we use only the
cleanest $D^0$ decay modes $D^0 \ra K^- \pi^+$ and $K^- K^+$. 
As in the $J/\psi\,D^{\ast+}$ study, we
remove $B\overline{B}$ events by requiring
$p^*_{D^0}>2.6\,\mathrm{GeV}/c$ or
$p^*_{\ell^\pm}>2.6\,\mathrm{GeV}/c$.

A plot of dilepton versus $K^- \pi^+(K^- K^+)$ masses,
and the projection onto the $K^- \pi^+(K^- K^+)$ mass axis,
are shown in Figs.~\ref{md}{\it c,d}.
A simultaneous fit to the $K^- \pi^+(K^- K^+)$ mass distribution in the
$J/\psi$ signal window and the sideband finds $N_{D^0}=15.9^{+5.4}_{-4.7}$ in 
the signal region, with a combinatorial $(\ell^+ \ell^-)\, D^0$ contribution
of $1.0 \pm 0.8$. The signal yield is $14.9^{+5.4}_{-4.8}$,
with significance $\sigma_{J/\psi\,D^0\,X} = 3.7$.
We use a Gaussian signal shape and a linear background function 
in the fit. As a cross-check, we also
fit the dilepton mass spectrum for $D^0$ signal and sideband regions,
obtaining $N_{J/\psi}=17.7^{+5.3}_{-4.6}$ and $N_{J/\psi}=4.3 \pm
0.8$ respectively, where the sideband number is the result of the fit
scaled to the expected contribution under the $D^0$ peak. For all
fits the signal shapes are fixed from the Monte Carlo simulation.

To study our reconstruction efficiency for $J/\psi$ mesons produced
together with a charmed meson, we generate $e^{+}e^{-}
\ra J/\psi\, c\bar{c}$ Monte Carlo events using a simple model adapted
from the {\tt QQ} event generator~\cite{qq}. Since the efficiency for
both particle reconstruction and selection criteria strongly depend on
the kinematics, we correct the general kinematic characteristics of
the Monte Carlo events to match those of the data using a
large sample of continuum $J/\psi$ events. In particular, the
distributions of the recoil mass, the $J/\psi$ production and
helicity angles, and the angle between the thrust axis of the
$c\bar{c}$ system and its boost are adjusted to match the data. These
quantities almost fully describe the kinematics of $e^{+}e^{-}
\ra J/\psi\, c\bar{c}$. The fragmentation function of $c\bar{c}$ into
charmed mesons in each bin of $Q^2(c\bar{c})$ is the only characteristic
that has an effect on the efficiency that cannot be determined from the data.
We vary the fragmentation function over a
wide range, taking the difference in the efficiency into account as
a systematic error.

The efficiency is first calculated for 
$p^{\ast}_{J/\psi} > 2.0\,\mathrm{GeV}/c$ and then extrapolated to the
full momentum interval, taking into account the cross-sections for
inclusive continuum production obtained in Ref.~\cite{behera} for both
$p^{\ast}_{J/\psi} > 2.0$ and  
$p^{\ast}_{J/\psi} < 2.0\,\mathrm{GeV}/c$ data.  The overall
efficiencies for $J/\psi\,D^{\ast +}$ and $J/\psi\,D^0$ are calculated to
be $\epsilon_{J/\psi\,D^{\ast+}}=(4.1 \pm 1.0) \cdot 10^{-4}$ and
$\epsilon_{J/\psi\,D^0}=(3.7 \pm 0.8)\cdot 10^{-4}$ respectively.
Using these values, we find cross-sections
$\sigma (e^{+}e^{-}\ra J/\psi\,D^{\ast+} X) =\rs\,{\mathrm{pb}}$ and
$\sigma (e^{+}e^{-} \ra J/\psi\,D^0 X)=\rd\,{\mathrm{pb}}$.
Contributions to the systematic
error are summarized in Table~\ref{sys2}.

According to the Lund model, $c\bar{c}$ fragmentation produces
charmed mesons at the rate of $0.53$ per event for $D^{\ast +}$,
and $1.18$ per event for $D^0$, where both 
numbers include feed-down from higher states 
(in particular, $D^{\ast +} \ra D^0 \pi^+$)~\cite{jetset}.
Assuming that these rates apply to $c\bar{c}$ fragmentation in
$e^{+}e^{-}\ra J/\psi\, c\bar{c}$, we calculate
$\sigma(e^{+}e^{-} \ra J/\psi\, c\bar{c})$ and find
\rscc$\,{\mathrm{pb}}$ and \rdcc$\,{\mathrm{pb}}$
based on our $D^{\ast+}$ and $D^0$ measurements,
respectively. No systematic error is included for
our use of the Lund fragmentation rates.
These results are slightly correlated, as two events are common
to both samples.  Taking this into account, we average the results 
and obtain \racc$\,{\mathrm{pb}}$.
In Ref.~\cite{behera} we
found the inclusive prompt $J/\psi$ cross-section to be
$\sigma(e^{+}e^{-} \ra J/\psi\,X) = (1.47 \pm 0.10 \pm 0.11)\,{\mathrm
{pb}}$, based on a $32.7\,{\rm {fb}}^{-1}$ dataset.
We therefore infer that a large fraction of prompt $J/\psi$
events,
$\sigma(e^{+}e^{-} \ra J/\psi\,c\bar{c}) /
 \sigma(e^{+}e^{-} \ra J/\psi\,X) = \raccrel$,
is due to the $e^{+}e^{-} \ra J/\psi\,c\bar{c}$ process.
Contributions to the systematic error on this ratio are shown in
Table~\ref{sys2} by the numbers in parentheses.

This $J/\psi\, c\bar{c}$ cross-section is
an order of magnitude larger than predicted in Refs.~\cite{nrqcd2a,nrqcd2d,lik},
and contradicts the NRQCD expectation that the $J/\psi\,c\bar{c}$
fraction is small~\cite{nrqcd2a,nrqcd2d}.
We note, however, that our result is dependent on the
fragmentation model assumed for the $c\bar{c}$ system.
In the future, more comprehensive measurements including associated
$J/\psi\,D^+$, $J/\psi\,D_s^+$ and $J/\psi\,\Lambda_c^+$
production could significantly reduce this model dependence.

\begin{table}[htb]
\caption{Systematic error contributions for
	$\sigma(e^+ e^- \to J/\psi\,D\,X$). Numbers in parentheses show
	contributions to the error on the ratio
	$\sigma(e^{+}e^{-} \ra J/\psi\,c\bar{c}) /
	\sigma(e^{+}e^{-} \ra J/\psi\,X)$.
 }
\label{sys2}
\begin{tabular}{@{\hspace{0.3cm}}l@{\hspace{0.3cm}}||
@{\hspace{0.3cm}}r@{\hspace{0.1cm}}r@{\hspace{0.3cm}}|
@{\hspace{0.3cm}}r@{\hspace{0.1cm}}r@{\hspace{0.3cm}}}
\hline \hline
	& \multicolumn{4}{c}{Systematic error (\%)}			\\ 
Source	& \multicolumn{2}{c}{$J/\psi\,D^0$}
	& \multicolumn{2}{c}{$J/\psi\,D^{\ast +}$}			\\
\hline
MC kinematics correction
	& $\pm 11$	& $(\pm 8)$	& $\pm 10$	& $(\pm 8)$	\\ 
$c\bar{c}$ fragmentation function
	& $\pm 8$	& $(\pm 8)$	& $\pm 15$	& $(\pm 15)$	\\ 
Fitting procedure
	& $\pm 10$	& $(\pm 10)$	& $\pm 5$	& $(\pm 5)$	\\ 
Efficiency of $p^{\ast}_{J/\psi}$ cut
	& $\pm 11$	& $(0)$		& $\pm 11$	& $(0)$		\\ 
Track reconstruction	
	& $\pm 8$	& $(\pm  4)$	& $\pm 12$	& $(\pm 8)$	\\
Lepton and $K$ identification
	& $\pm 6$	& $(\pm  3)$	& $\pm 6$	& $(\pm 3)$	\\
\hline
Total	& $23$		& $(16)$	& $26$		& $(20)$	\\
\hline\hline
\end{tabular}
\end{table}

In summary, we have observed both a charmonium state and charmed
mesons accompanying prompt $J/\psi$ production in $e^+ e^-$
annihilation.  We measure 
$\sigma (e^{+}e^{-} \ra J/\psi\, \eta_c (\gamma)) \times
{\mathcal{B}}(\eta_c \ra \,\geq 4\;{\rm charged})=\re\,{\mathrm{pb}}$
and $\sigma (e^{+}e^{-} \ra J/\psi\,D^{\ast +}
X)=\rs\,{\mathrm{pb}}$, 
and estimate
$\sigma(e^{+}e^{-} \ra J/\psi\,c\bar{c}) /
 \sigma(e^{+}e^{-} \ra J/\psi\,X) = \raccrel$.
Our results favor $e^{+}e^{-} \ra J/\psi\,c\bar{c}$ as the leading
mechanism for prompt $J/\psi$ production at
$\sqrt{s} \approx 10.6\,\mathrm{GeV}$.

We wish to thank the KEKB accelerator group for the excellent
operation of the KEKB accelerator.
We acknowledge support from the Ministry of Education,
Culture, Sports, Science, and Technology of Japan
and the Japan Society for the Promotion of Science;
the Australian Research Council
and the Australian Department of Industry, Science and Resources;
the National Science Foundation of China under contract No.~10175071;
the Department of Science and Technology of India;
the BK21 program of the Ministry of Education of Korea
and the CHEP SRC program of the Korea Science and Engineering Foundation;
the Polish State Committee for Scientific Research
under contract No.~2P03B 17017;
the Ministry of Science and Technology of the Russian Federation;
the Ministry of Education, Science and Sport of the Republic of Slovenia;
the National Science Council and the Ministry of Education of Taiwan;
and the U.S.\ Department of Energy.


\begin{thebibliography} {99}

\bibitem{behera} K.~Abe {\it et al.} (Belle Collab.),
Phys. Rev. Lett. {\bf 88}, 052001 (2002).

%
%
\bibitem{nrqcd1}
E.~Braaten and S.~Fleming, Phys. Rev. Lett. {\bf 74}, 3327 (1995);
P.~Cho and M.~Wise, Phys. Lett. {\bf B346}, 129 (1995);
M.~Cacciari, M.~Greco, M.~L.~Mangano, and A.~Petrelli,
\emph{ibid.}\ {\bf B356}, 553 (1995).

%
%
\bibitem{nrqcd2a}
P.~Cho and A.~K.~Leibovich, Phys. Rev. D {\bf 53}, 150 (1996);
{\bf 53}, 6203 (1996).
S.~Baek, P.~Ko, J.~Lee, and H.~S.~Song, J. Kor. Phys. Soc. {\bf 33}, 97 (1998);
hep-ph/9804455.

%
%
\bibitem{nrqcd2c}
E.~Braaten and Yu-Qi Chen, Phys. Rev. Lett. {\bf 76}, 730 (1996).

%
%
\bibitem{nrqcd2d}
F.~Yuan, C.-F.~Qiao, and K.-T.~Chao, Phys. Rev. D {\bf 56}, 321 (1997).

\bibitem{babar} B.~Aubert {\it et al.} (BaBar Collab.),
Phys. Rev. Lett. {\bf 87} 162002 (2001).

\bibitem{iwa} Y.~Iwasaki,  Phys. Rev. D {\bf 16}, 220 (1977) .

\bibitem{kane} G.~L.~Kane, J.~P.~Leveille, and  D.~M.~Scott,
Phys. Lett. {\bf B85}, 115 (1979). 

\bibitem{lik} V.~V.~Kiselev, A.~K.~Likhoded, and M.~V.~Shevlyagin,
Phys. Lett. {\bf B332}, 411 (1994).

\bibitem{KEKB} KEKB B Factory Design Report, KEK Report 95-7, 1995
(unpublished); Y.~Funakoshi {\it et al.}, Proc. 2000 European Particle
Accelerator Conference, Vienna 2000.

\bibitem{Belle} A.~Abashian {\it et al.} (Belle Collab.), 
Nucl. Instr. and Meth. {\bf A479}, 117 (2002). 

\bibitem{etaprimemass} 
W.~Buchm\"{u}ller and S.-H.~H.~Tye, Phys. Rev. D. {\bf 24}, 132 (1981);
S.~Godfrey and N.~Isgur, \emph{ibid.}\ {\bf 32}, 189 (1985).

\bibitem{PDG}
D.E.~Groom {\it et al.}, Eur. Phys. J. {\bf C15}, 1 (2000).

\bibitem{implied}
Charge-conjugated modes are implicitly included.

\bibitem{qq}
{\tt QQ} was developed by the CLEO Collaboration. See
http://www.lns.cornell.edu/public/CLEO/soft/QQ.

\bibitem{jetset}
T. Sj\"ostrand, \textsc{Pythia} 5.7 / \textsc{Jetset} 7.4,
Comp. Phys. Commun. {\bf 82}, 74 (1994).
The $D^{\ast +}/D^0$ ratio is constant over $\sqrt{s}$ except for a decrease
near $D^{\ast +}$ mass threshold: the effect on the result is negligible.


\end{thebibliography}
\end{document}